# Virtual Social Immersive Multi-Sensory E-Commerce

Alpana Dubey[*]  
Accenture Labs, India

Suma Mani Kuriakose[!]  
Accenture Labs, India

Sumukha Anand[‡]  
Accenture Labs, India

Nitish Bhardwaj[§]  
Accenture Labs, India

Shubhashis Sengupta[¶]  
Accenture Labs, India

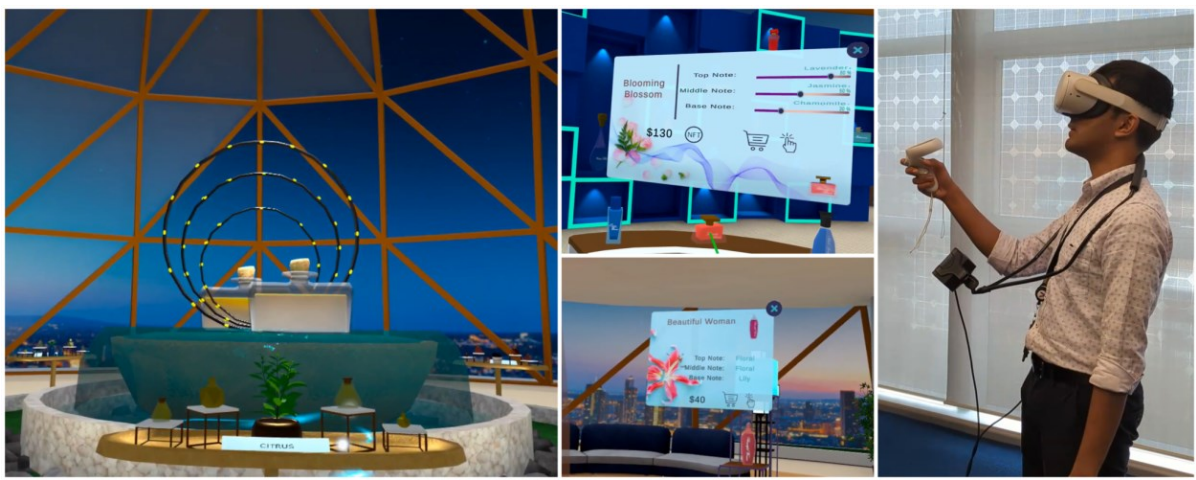

Figure 1: Aromaverse VR application showcasing customizable and off-the-shelf perfumes.

**ABSTRACT**

In this paper, we present a virtual immersive multi-sensorial experience, Aromaverse. Aromaverse is an immersive 3D multi-player environment augmented with olfactive experience where users can experience and customize perfumes. Being multi-player, users can join the same space and enjoy a social buying experience. The olfactive experience embodied in the perfume allows users to experience their fragrances. This further enhances the user perception of perfumes in a virtual setting. Aromaverse also provides the ability to customize the perfumes by changing their top, mid, and base notes. The customized fragrances can be shared with other users, enabling a shared olfactive experience. To understand users' buying experience in such an environment, we conducted a set of experiments in which participants were requested to explore the space, experience the perfumes, customize them and buy them. They were asked to perform the same activities alone and in the presence of their friends. Various factors including the benefits and limitations of such an experience were captured by the questionnaires. Our results show that the presence of a companion enhances the shopping experience by improving the level of imagination of the product and helping in making purchase decisions. Our findings suggest that multi-sensorial XR experiences offer great opportunities to retail firms to improve customer engagement and provide more realistic online experience of products that require other sensory modalities.

## 1 INTRODUCTION

The use of online shopping platforms has increased significantly in recent times primarily due to the improvement in supply chains, inventory, and warehouse management. Although online shopping applications, such as web-based or mobile apps, offer convenience to their users, they lack the immersive and spatial experience that physical stores offer. The spatial and social experience of physical stores gives shoppers a multi-sensorial medium to experience products. Physical stores can arrange products in an aesthetically pleasing spatial environment which helps to drawing customers' attention and enhance sales.

Additionally, physical stores allow shopping with a companion which is more enjoyable and satisfying as the companion can provide moral support and opinions on the product [1]. Some individuals may prefer online shopping for its convenience, while others may prefer a physical store for better socialization. Studies indicate that customers seek both utilitarian (practical) and hedonic (enjoyable) purchasing experiences. The technologies that enhance these aspects can greatly improve consumer satisfaction [2]. Multi-Sensory experiences in VR provide highly immersive experiences. These multi-sensory components must be congruent with other components of the application for it to be most effective, failure to maintain this congruency will impact and disrupt the overall experience [3].

Virtual Reality (VR)/ Mixed Reality (MR) / Extended Reality (XR) experiences can offer the benefits of physical store shopping without compromising the convenience of online shopping. However, the present XR experiences lack multi-sensory experiences that we often take for granted in the physical stores. As the perception of several products is shaped through sensory modalities beyond audio-visual experiences, a multi-sensory experience is highly desirable for such products [4]. We propose Aromaverse - a virtual immersive multi-sensorial experience for fragrance retail, where users can experience and buy perfumes. Aromaverse leverages olfactory simulations and allows users to engage with perfumes multi-sensorially, i.e. by interacting with perfume bottles arranged in a spatial environment and experiencing the fragrances. Being multi-player, Aromaverse offers a social shopping experience. The platform also supports product fragrance customization allowing users to change the top, mid and base notes. We conducted a user study to address the following research questions on the effect of multisensorial VR environments on consumer's purchase experiences: 1) How does the presence of a companion impact the multi-sensory VR shopping experience? 2) How does the sense of presence and realism corelate with user engagement and immersion? Our results show that such environments help in increased immersion and higher buying satisfaction.

*e-mail: alpana.a.dubey@accenture.com  
!e-mail: suma.mani.kuriakose@accenture.com  
‡e-mail: sumukha.anand@accenture.com  
§e-mail: nitish.a.bhardwaj@accenture.com  
¶e-mail: shubhashis.sengupta@accenture.com

## 2 AROMAVERSE

Aromaverse is a multi-player environment where users can choose their avatars and usernames. It is integrated with a sophisticated aroma emitting device. The application comprises of different zones where players can choose to experience perfumes. A perfume comprises of three notes: top, mid and base notes. The application features two types of perfumes: customized and off-the-shelf. In the customized perfumes the player is given an option to change the notes. Conversely, the notes in the off-the-shelf perfumes can't be changed.

The application can be deployed on a standalone VR headset. For navigation in the space, the player can use teleportation or controller joysticks operated by their thumbs. Communication with other players was facilitated through voice chat, which is enabled by default.

## 3 EXPERIMENT AND RESULTS

The objective of the study is to understand the impact of shopping alone vs. with a companion in a VR environment and the associated user satisfaction. The user study group consisted of 13 participants: 7 male and 6 female participants belonging to diverse age groups and familiarity with VR.

Each participant was required to shop for perfumes in the environment with an allotted budget of $300. A participant would make two trips -1) alone and 2) with a companion of their choice. Conversely, if the featured perfumes were not up to their liking, they could customize the perfumes to create a signature scent; these perfumes were priced at a premium as compared to off-the-shelf ones. Customized perfumes were set at the price range of $120 to $130, whereas the off-the-shelf perfumes ranged from $30 to $50. Time spent in the experience and number of perfumes bought along with the total amount spent were recorded.

Prior to the experiment, participants were given a training on how to navigate and buy products in the VR application and post the experiment they were requested to complete a questionnaire which included a total of 14 questions including 3 open ended questions. The questionnaire aimed to measure the user's overall experience and satisfaction, immersion, usability, influence on purchase decision, spending behavior and comparison with traditional shopping. We found that the presence of a companion positively increased the experience time, level of immersion, imagination of the scent, purchase experience, and navigation (Figure 2).

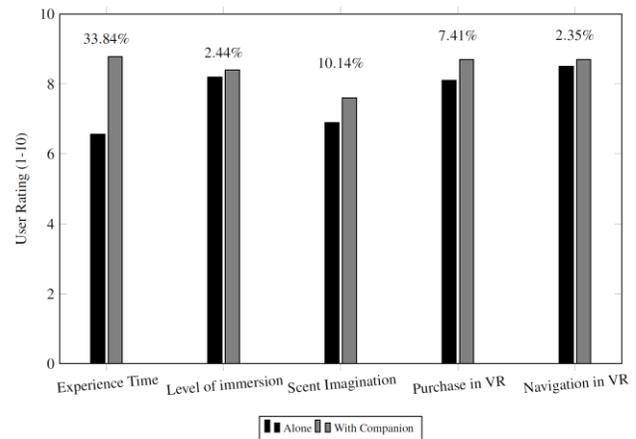

Figure: 2 Visualization of aggregated responses of all participants across different elements.

The responses to open-ended questions revealed both the advantages and the limitations of such an experience. The innovative experience of digital purchasing with virtually sampled fragrances was the highlight for the participants. Participants enjoyed shopping with a companion in an uncrowded virtual space. The customization of perfumes and the virtual in-store experience were also appreciated. The most reported concerns were the precision of aroma trials, discomfort of wearing VR headsets and challenges associated with navigation within the VR application. Participants found the VR shopping a compelling combination of online and physical store shopping. Few participants still hold a strong preference for the touch sensation, ambience and familiarity of physical stores especially when buying products such as perfumes.

## 4 CONCLUSION

In this paper, we present Aromaverse, an immersive social multi-sensory perfume buying experience. While Aromaverse shows great potential, especially in the ability to offer novel and immersive ways to shop, addressing challenges related to comfort especially in long term use and seamless integration of multi-sensory elements into the application will be crucial for wider adoption. Further research should focus on enhancing the accuracy of such multi-sensory devices in comparison with the actual product to enhance overall user satisfaction and trust.